\documentclass[preprint,showpacs,eqsecnum,amsmath,amssymb,nofootinbib]{revtex4}

\usepackage{graphicx}

\usepackage{amssymb,latexsym}
\usepackage{epstopdf}

\newcommand{\bv}{\bar{\varphi}}
\newcommand{\rot}{\widetilde{\rho}}

\begin{document}

\title{Renormalization and vacuum energy for an interacting scalar field in a $\delta$-function potential}

\author{David J. Toms}
\homepage{http://www.staff.ncl.ac.uk/d.j.toms}
\email{d.j.toms@newcastle.ac.uk}
\affiliation{%
School of Mathematics and Statistics,
Newcastle University,
Newcastle upon Tyne, U.K. NE1 7RU}

\date{\today}

\begin{abstract}
We study a self-interacting scalar field theory in the presence of a $\delta$-function background potential. The role of surface interactions in obtaining a renormalizable theory is stressed and demonstrated by a two-loop calculation. The necessary counterterms are evaluated by adopting dimensional regularization and the background field method. We also calculate the effective potential for a complex scalar field in a non-simply connected spacetime in the presence of a $\delta$-function potential. The effective potential is evaluated as a function of an arbitrary phase factor associated with the choice of boundary conditions in the non-simply connected spacetime. We obtain asymptotic expansions of the results for both large and small $\delta$-function strengths, and stress how the non-analytic nature of the small strength result vitiates any analysis based on standard weak field perturbation theory.
\end{abstract}


\maketitle

\section{\label{intro} Introduction}
\setcounter{equation}{0}
\subsection{Dedication}

It is a great pleasure to contribute this article to honour Stuart Dowker's many achievements in theoretical physics. I first knew Stuart by reputation long before I ever had the chance to meet him. When I was a postgraduate student my main interest was in quantum field theory in curved spacetime. Long before the internet and the arXiv preprints in high and theoretical energy physics used to be sent by post to SLAC where they would appear on a list that was sent out weekly to subscribers. At that time there were four main groups in the UK working on quantum field theory in curved spacetime: the Cambridge group, based around Stephen Hawking and Gary Gibbons; the King's College group based around Paul Davies; the Imperial College group, based around Chris Isham and Mike Duff; and the Manchester group, consisting of Stuart Dowker and his students. The SLAC preprint list would be scanned avidly for anything coming from these groups, but as I was especially interested in quantum field theory in topologically non-trivial spacetimes and vacuum energy calculations using $\zeta$-function methods, the work of Stuart Dowker was always eagerly awaited. An especially treasured preprint, that I still use, was {\it Selected Topics in Topology and Quantum Field Theory\/} that was based on lectures given by Stuart at Austin, and which were never published; these notes pre-date all of the now standard books and reviews on topology and gauge fields.

Since coming to the UK, Stuart and I have met on many occasions, often at Ph.~D. vivas. I am sure that many contributors to this volume will have had the ``Dowker experience'': You mention to Stuart that you are working on some calculation to be met with something like ``Oh yes. That is just a special case of a theorem by Schnekelgreuber\footnote{So far as I am aware I made this name up, but I would not be at all surprised if Stuart was to say Scnekelgreuber really existed but was only active between 1829 and 1837, not in 1878.} published in 1878.''

\subsection{Background}
The problem of computing the vacuum, or Casimir, energies in the case of non-smooth background potentials has a long history, and in addition has become the focus of much recent interest. Of especial interest to the present paper is the case of $\delta$-function background potentials.  One of the pioneering, and most important, papers on the calculation of vacuum energies in $\delta$-function potentials is \cite{BHR} whose analysis is based on earlier Green function calculations of \cite{HR}. Later work on vacuum energy and other related quantum field theory calculations includes \cite{Solod,Scad1,Scad2,Grahametal0,Grahametal1,MiltonPRD,Grahametal2,MiltonJPA37,BordagVass,Jaffe,TomsPLBdelta,Khus}. Some motivation for these investigations can be garnered from the study of brane-world models stimulated by the Randall-Sundrum scenario \cite{RS} where non-smooth backgrounds in the form of $\delta$-function potentials arise in a natural manner that is crucial to the features of these models. There are many studies of vacuum energies in brane-world spacetimes. (See \cite{Garriga,TomsRS,GRoth,FlachiToms,Saharian,KnapToms} for some of the early investigations.)

Almost all of the previous studies of vacuum energies in non-smooth backgrounds are concerned with non-interacting (apart from interactions with the background) fields. This renders the renormalization considerations somewhat simpler than is the case when interacting fields are involved, particularly beyond one-loop order. The renormalization of $\lambda\phi^4$ theory in a $\delta$-function background was considered by \cite{BordagVass} but at one-loop order only where additional divergences were dealt with by the addition of extra surface terms to the original action. A different interpretation of this procedure was given by \cite{TomsPLBdelta} (see also the later paper \cite{Khus}) that is more in keeping with standard renormalization theory, and is the method that we will adopt in this paper, to be detailed in Sec.~\ref{renorm} below.

One of the key features that we must deal with when proceeding beyond one-loop order is the presence of overlapping divergences, and the necessity of showing that they cancel (otherwise there will be non-local divergences present that cannot be removed by local counterterms). We will adopt the background field method and dimensional regularization as in \cite{DJTrenorm}. A comprehensive review of the method for smooth backgrounds can be found in \cite{ParkerTomsbook}. The general issue of interacting fields on manifolds with boundaries, smooth or not, beyond one-loop order appears to have received little attention. (The case of smooth boundaries at one-loop order is dealt with in \cite{DJTrenorm}. Tsoupros \cite{Tsoupros1,Tsoupros2,Tsoupros3} examines smooth spherical cap geometries and Haba \cite{Haba} considers $\delta$-function backgrounds. The calculations presented below do not overlap directly with these references.)

The outline of our paper is the following. In Sec.~\ref{renorm} we discuss the renormalization of an interacting scalar field theory with cubic and quartic self-interactions to two-loop order in a $\delta$-function potential. Dimensional regularization is used. The background spacetime is flat and assumed to be 4-dimensional with a possible periodic identification of one spatial coordinate. The importance of including correct boundary terms, especially a boundary interaction in the field (see (\ref{R3}) below), is stressed. Without the proper boundary terms the theory will not be renormalizable. Complications in the evaluation of the complete two-loop effective action are highlighted, and it is shown how to renormalize the effective potential to two-loop order. Necessary two-loop counterterms are found. In Sec.~\ref{vacuum} we address the calculation of the one-loop effective potential for a complex scalar field in a $\delta$-function potential with one spatial coordinate periodically identified to form a circle. This allows the complex field to change by an arbitrary phase around the circle and we calculate the effective potential as a function of this phase. A number of asymptotic limits are obtained. Sec.~\ref{discuss} summarizes our results and presents a short discussion. Appendix A outlines a method for evaluating the Green function in the case of any number of $\delta$-function potentials. Appendix B describes some technical details in the evaluation of certain types of integrals that arise in the renormalization beyond one-loop order.

\section{\label{renorm} Renormalization}
\setcounter{equation}{0}

We consider an interacting scalar field in four spacetime
dimensions with a general self-interaction. The theory will be
regularized using dimensional regularization as in
\cite{DJTrenorm}. The bare action will be chosen to be
\begin{equation}\label{R1}
S=S_{\rm bulk}+S_{\rm bdry}\;,
\end{equation}
where
\begin{eqnarray}\label{R2}
S_{\rm bulk}&=&\int d^{D+1}x\Big\lbrace\frac{1}{2}
\partial^\mu\phi_B\partial_\mu\phi_B+\frac{1}{2}m_B^2\phi_B^2+\Lambda_B\nonumber\\
&&+h_B\phi_B+\frac{1}{3!}g_B\phi_B^3+\frac{1}{4!}\lambda_B\phi_B^4\Big\rbrace\;,
\end{eqnarray}
and
\begin{equation}\label{R3}
S_{\rm bdry}=\int\limits_{y=a}d^Dx\Big\lbrace\frac{1}{2}v_B\phi_B^2+\gamma_B\phi_B+\sigma_B\Big\rbrace\;.
\end{equation}
The bulk action $S_{\rm bulk}$ extends over the complete $(D+1)$
dimensional spacetime. The boundary term $S_{\rm bdry}$ only extends
over the $D$-dimensional subspace specified by $y=a$ that gives
the location of the $\delta$-function. $S_{\rm bdry}$ could obviously be written in the equivalent form of an integral over the $(D+1)$-dimensional spacetime with the terms in braces in (\ref{R3}) multiplying the $\delta$-function $\delta(y-a)$. All of the coupling
constants that occur in the action are bare unrenormalized
expressions. We might have expected, on dimensional grounds, that
a $\phi_B^3$ term appeared in $S_{\rm bdry}$ since its coefficient
would be dimensionless for the case of interest $D=3$. However,
this means that the associated $\phi_B^3$ counterterm could not
depend on the strength of the $\delta$-function (that has units of
mass, or inverse length), and because in the absence of a
$\delta$-function potential there is no need for $S_{\rm bdry}$, we
rule out such a term. We expect that $S_{\rm bdry}\rightarrow0$ as
$v\rightarrow0$, where $v$ is the renormalized $\delta$-function
strength. (This conclusion is substantiated in the explicit calculation presented below.)

In dimensional regularization we require~\cite{tHooft4} all renormalized
couplings to have the same dimensions for all $D$ as they do when
$D=3$. We write
\begin{equation}\label{epsilon}
D=3+\epsilon\;,
\end{equation}
and introduce an arbitrary unit of length $\ell$. (It is more customary \cite{tHooft4} to use a unit of mass $\mu$, but obviously $\ell=1/\mu$, and we stick with $\ell$.) Bare quantities
are expressed in terms of renormalized ones by
\begin{eqnarray}
\phi_B&=&\ell^{-\epsilon/2}Z^{1/2}\phi\;,\label{R4a}\\
m_B^2&=&m^2+\delta m^2\;,\label{R4b}\\
\Lambda_B&=&\ell^{-\epsilon}(\Lambda+\delta\Lambda)\;,\label{R4c}\\
h_B&=&\ell^{-\epsilon/2}(h+\delta h)\;,\label{R4d}\\
g_B&=&\ell^{\epsilon/2}(g+\delta g)\;,\label{R4e}\\
\lambda_B&=&\ell^{\epsilon}(\lambda+\delta \lambda)\;,\label{R4f}\\
v_B&=&v+\delta v\;,\label{R4g}\\
\gamma_B&=&\ell^{-\epsilon/2}(\gamma+\delta \gamma)\;,\label{R4h}\\
\sigma_B&=&\ell^{-\epsilon}(\sigma+\delta\sigma)\;.\label{R4i}
\end{eqnarray}
We write the field renormalization constant $Z$ as
\begin{equation}\label{R4j}
Z=1+\delta Z\;,
\end{equation}
and all counterterms are expanded in powers of $\hbar$, the loop
counting parameter:
\begin{equation}\label{R5}
\delta C=\hbar\delta C^{(1)}+\hbar^2\delta C^{(2)}+\cdots\;.\;,
\end{equation}
Here $C$ denotes a generic quantity that occurs in
(\ref{R4b}--\ref{R4j}). The counterterms $\delta C^{(1)},\delta
C^{(2)},\ldots$ are chosen to contain pole terms in $\epsilon$
that cancel the various poles arising in the effective action
order by order in the loop expansion.

The effective action can be evaluated in the loop expansion in a
suitable form for a discussion of renormalization as described by
Jackiw~\cite{Jackiw}. Here we follow the formalism used to
consider renormalization in curved
spacetime~\cite{DJTrenorm,HuishToms}. The only added feature here
is the presence of $S_{\rm bdry}$. This affects the Green function, as
well as the interaction vertices beyond that occurring for the
bulk theory. The basic step is to write
\begin{equation}\label{background}
\phi=\bv+\hbar^{1/2}\psi\;,
\end{equation}
with $\bv$ the background field. The results
(\ref{R4a}--\ref{R4j}) are used in (\ref{R2}) and (\ref{R3}), all
counterterms are expressed as in (\ref{R5}) and the result
expanded out to a consistent order in $\hbar$. Standard functional
methods then generate the loop expansion of the effective action.
The result can be expressed as
\begin{equation}\label{R6}
\Gamma\lbrack\bv\rbrack=S\lbrack\bv\rbrack+\frac{\hbar}{2}{\rm tr}\ln(\ell^2\bar{\Delta})+\hbar \left\langle 1-\exp\left(-\frac{1}{\hbar}S_{\rm int}\right)\right\rangle\;,
\end{equation}
where $\bar{\Delta}$ is the operator that defines the Green
function following from the second functional derivative of the
action functional with respect to the field evaluated at the
background field. (See (\ref{R7}) and (\ref{R8}) below.) The terms
of cubic and higher orders in the field, as well as the
counterterms, are defined as the interaction part of the action
$S_{\rm int}$ and treated perturbatively. The angular brackets in
(\ref{R6}) denote an evaluation using Wick's theorem with only
one-particle irreducible graphs kept. We define
\begin{equation}\label{R7}
\langle\phi(x)\phi(x')\rangle=G_v(x,x')\;,
\end{equation}
where
\begin{equation}\label{R8}
\left\lbrack-\Box_x+m^2+g\bv
+\frac{\lambda}{2}\bv^2+v\delta(y-a)\right\rbrack G_v(x,x')
=\delta(x,x')\;.
\end{equation}

We can express $S_{\rm int}$ in terms of $\hbar$ by defining
\begin{equation}\label{R9}
\frac{1}{\hbar}S_{\rm int}=\hbar^{1/2}A_1\lbrack\bv,\psi\rbrack +\hbar
A_2\lbrack\bv,\psi\rbrack+{\mathcal O}(\hbar^{3/2})\;.
\end{equation}
In our case,
\begin{eqnarray}
A_1&=&\ell^{-\epsilon}\int d^{D+1}x\;\frac{1}{3!}\left\lbrace g+\lambda\bv(x)\right\rbrace\psi^3(x)\;,\label{R10a}\\
A_2&=&\ell^{-\epsilon}\int d^{D+1}x\left\lbrace\frac{\lambda}{4!}\psi^4+\frac{1}{2}\delta m^{2(1)}\psi^2+\frac{1}{2}\delta g^{(1)}\bv(x)\psi^2 +\frac{1}{4!}\delta\lambda^{(1)}\bv^2(x)\psi^2\right\rbrace\nonumber\\
&&+\ell^{-\epsilon}\int\limits_{y=a}d^Dx\;\frac{1}{2}\delta v^{(1)}\psi^2\;.\label{R10b}
\end{eqnarray}
The only difference with what happens in the absence of a
$\delta$-function potential is the presence of the last term in
$A_2$, and an alteration of the Green function. (Of course the
alteration of the Green function affects the evaluation of the
effective action considerably.)

Expanding the effective action to two-loop order results in (see \cite{TomsSAP} for a review)
\begin{eqnarray}
\Gamma\lbrack\bv\rbrack&=&S\lbrack\bv\rbrack
+\frac{\hbar}{2}{\rm tr}\ln(\ell^2\bar{\Delta})
+\hbar^2\left(\langle A_2\rangle-
\frac{1}{2}\langle A_1^2\rangle\right)+\cdots\nonumber\\
&=&S\lbrack\bv\rbrack +\frac{\hbar}{2}{\rm tr}
\ln\left(\ell^2\bar{\Delta}\right)+\hbar^2\Gamma^{(2)}+\cdots\;,\label{R11}
\end{eqnarray}
where
\begin{eqnarray}
\Gamma^{(2)}&=&\ell^{-\epsilon}\int
d^{D+1}x\Big\lbrace\frac{\lambda}{8}G_v^2(x,x)+
\frac{1}{2}\delta m^{2(1)}G_v(x,x)\nonumber\\
&+&\frac{1}{2}\delta g^{(1)}\bv(x)G_v(x,x)+
\frac{1}{4}\delta\lambda^{(1)}\bv^2(x)G_v(x,x)\Big\rbrace\nonumber\\
&+&\ell^{-\epsilon}\int\limits_{y=a}d^Dx\;
\frac{1}{2}\delta v^{(1)}G_v(x,x)\nonumber\\
&-&\frac{1}{12}\ell^{-2\epsilon}\int
d^{D+1}xd^{D+1}x'\lbrack g+\lambda\bv(x)\rbrack \lbrack g+\lambda\bv(x')\rbrack G_v^3(x,x')\;,\label{R12}
\end{eqnarray}
gives the two-loop contribution to the effective action. The next two subsections will examine the one- and two-loop divergences of the effective action that we have obtained.

\subsection{\label{oneloop}One-loop effective action}

The one-loop part of the effective action ({\ref{R11}) involves
\begin{equation}\label{R13}
\Gamma^{(1)}=\frac{1}{2}{\rm
tr}\ln\left(\ell^2\bar{\Delta}\right)\;.
\end{equation}
In order to deal with this expression using dimensional
regularization, we can differentiate with respect to $m^2$ and use
the fact that the inverse of $\bar{\Delta}$ is the Green function
as defined by (\ref{R8}). We find
\begin{equation}\label{R14}
\frac{\partial}{\partial m^2} \Gamma^{(1)}=\frac{1}{2}\int
d^{D+1}x\;G_v(x,x)\;.
\end{equation}
It is sufficient to compute $G_v(x,x)$ with constant background
fields, since using the derivative expansion and power counting
shows that there can be no pole terms that involve derivatives of
the background field at one-loop order. (This reflects the
well-known result that field renormalization is first required at
two-loop order. See \cite{TomsSAP} for a proof of this using the derivative expansion of the effective action.) We can then use the result in (\ref{A3}) with
$m^2$ replaced by
\begin{equation}\label{R15}
M^2=m^2+g\bv+\frac{\lambda}{2}\bv^2\;.
\end{equation}
$\bv$ is now regarded as constant. Use of (\ref{A3}) gives us
\begin{equation}\label{R16}
\frac{\partial}{\partial m^2} \Gamma^{(1)}=\frac{1}{2}V_D\int
\limits_{-L/2}^{L/2}dy\int \frac{d^Dp}{(2\pi)^D} \;G_v(p;y,y)\;,
\end{equation}
where $V_D=\int d^Dx$ is the volume associated with
$\mathbf{x}_\perp$. If we use the result in (\ref{A16}) the
integration over $y$ may be performed with the result
\begin{eqnarray}
\frac{\partial}{\partial m^2} \Gamma^{(1)} &=&\frac{1}{2}V_D\int
\frac{d^Dp}{(2\pi)^D} \left\lbrace
\frac{L}{2\omega_p}\coth(L\omega_p/2)\right.\nonumber\\
&&\left.-\frac{v\left\lbrack L\omega_p+\sinh(L\omega_p)
\right\rbrack}{8\omega_p^3\sinh^2(L\omega_p/2) \left\lbrack 1+
\frac{v}{2\omega_p}\coth(L\omega_p/2)\right\rbrack}\right\rbrace
\;.\label{R17}
\end{eqnarray}
Note that because of the replacement of $m^2$ with $M^2$ in
(\ref{R15}), we should use $\omega_p=\left(p^2+M^2 \right)^{1/2}$
here.

If we are only interested in the pole part of $\Gamma^{(1)}$ to
discuss the renormalization, then we may study the large $p$
behaviour of the integrand. If we denote the pole part of
$\Gamma^{(1)}$ by ${\rm PP}(\Gamma^{(1)})$, then it follows that
\begin{eqnarray}\label{R18}
\frac{\partial}{\partial m^2}{\rm PP}\left(
\Gamma^{(1)}\right)&=&\frac{1}{2}V_D\ {\rm PP}\left(\int
 \frac{d^Dp}{(2\pi)^D}\left\lbrack\frac{L}{2\omega_p}\right.\right.\nonumber\\
&&\hspace{25mm}\left.\left. -\frac{v}{4\omega_p^3}\left(1+\frac{v}{2\omega_p}\right)^{-1}\right\rbrack\right)\;.
\end{eqnarray}
Terms that have been dropped here decay exponentially for large
$p$. Although they contribute to the finite part of
$\Gamma^{(1)}$, they make no contribution to the poles. Using the
definitions in (\ref{B1}) and (\ref{B5}), and using the recursion
relation (\ref{B4}) shows that
\begin{eqnarray}\label{R19}
\frac{\partial}{\partial m^2}{\rm PP}\left(
\Gamma^{(1)}\right)&=&\frac{1}{2}V_D\ {\rm
PP}\left(\frac{L}{2}I(1)-\frac{v}{4}I(3)\right.\nonumber\\
&&\hspace{15mm}\left.+\frac{v^2}{8}I(4)-\frac{v^3}{16}I(5)+\cdots\right)\;.
\end{eqnarray}
Because of the simple dependence of $M^2$ on $m^2$, using
(\ref{B2}) results in
\begin{eqnarray}
{\rm PP}\left( \Gamma^{(1)}\right)&=&-\frac{1}{2}V_D(4\pi)^{-D/2}
\ {\rm PP}\Big\lbrace
\frac{L}{2}\frac{\Gamma(-1/2-D/2)}{\Gamma(1/2)}(M^2)^{D/2+1/2}\nonumber\\
&&-\frac{v}{4}\frac{\Gamma(1/2-D/2)}{\Gamma(3/2)}(M^2)^{D/2-1/2}
+\frac{v^2}{8}\frac{\Gamma(1-D/2)}{\Gamma(2)}(M^2)^{D/2-1}\nonumber\\
&&-\frac{v^3}{16}\frac{\Gamma(3/2-D/2)}{\Gamma(5/2)}(M^2)^{D/2-3/2}+\cdots\Big\rbrace\;.\label{R20}
\end{eqnarray}
The next term to that indicated will be finite as $D\rightarrow3$
and therefore contains no pole. By taking $D=3+\epsilon$ and
expanding about $\epsilon=0$ we find
\begin{equation}\label{R21}
{\rm PP}\left(
\Gamma^{(1)}\right)=\frac{V_3}{96\pi^2\epsilon}\left(
3LM^4+6vM^2-v^3\right)\;.
\end{equation}

Given the form of the counterterms in ({\ref{R4a}--\ref{R5}), with
$\delta Z^{(1)}=0$ and $\bv$ constant, we have the one-loop
counterterm part of the action as
\begin{eqnarray}
S_{\rm ct}&=&\hbar V_3L\Big\lbrace \frac{1}{2}\delta
m^{2(1)}\bv^2+\delta\Lambda^{(1)}+\delta h^{(1)}\bv\nonumber\\
&&+\frac{1}{3!}\delta
g^{(1)}\bv^3+\frac{1}{4!}\delta\lambda^{(1)}\bv^4\Big\rbrace\nonumber\\
&&+\hbar V_3\Big\lbrace \frac{1}{2}\delta
v^{(1)}\bv^2+\delta\gamma^{(1)}\bv+\delta\sigma^{(1)}
\Big\rbrace\;.\label{R22}
\end{eqnarray}
Note that $\int d^4x=V_3L$ identifies the bulk part of the action
and the $L$-independent part identifies the boundary part. Using
the expression for $M^2$ in (\ref{R15}) in the pole part of the
one-loop effective action in (\ref{R21}), it is easy to show that
the resulting effective action is finite as $\epsilon\rightarrow0$
if we choose the counterterms to be
\begin{eqnarray}
\delta\Lambda^{(1)}&=&-\frac{m^4}{32\pi^2\epsilon}\;,\label{R23a}\\
\delta h^{(1)}&=&-\frac{gm^2}{16\pi^2\epsilon}\;,\label{R23b}\\
\delta m^{2(1)}&=&-\frac{1}{16\pi^2\epsilon}(\lambda m^2+g^2)\;,\label{R23c}\\
\delta g^{(1)}&=&-\frac{3\lambda g}{16\pi^2\epsilon}\;,\label{R23d}\\
\delta \lambda^{(1)}&=&-\frac{3\lambda^2}{16\pi^2\epsilon}\;,\label{R23e}\\
\delta v^{(1)}&=&-\frac{\lambda v}{16\pi^2\epsilon}\;,\label{R23f}\\
\delta \gamma^{(1)}&=&-\frac{vg}{16\pi^2\epsilon}\;,\label{R23g}\\
\delta
\sigma^{(1)}&=&-\frac{v}{16\pi^2\epsilon}(m^2-\frac{1}{6}v^2)\;.\label{R23h}
\end{eqnarray}
The counterterms that enter the boundary part of the action all
vanish as $v\rightarrow0$ as would be expected. The counterterms
that enter the bulk part of the action agree with the more general
case considered in Ref.~\cite{DJTrenorm}.

\subsection{\label{twoloop}Two-loop effective action}

The complete two-loop effective action was given in (\ref{R12}).
It has not been possible to evaluate even the pole part of this
expression for the general scalar field theory that we have been
considering. The difficulty with performing a
complete renormalization calculation concerns the extraction of
the complete pole part of $G_v^3(x,x')$, a problem that we were
not able to overcome due to the complicated nature of the Green
function. Instead we will examine the simpler, but still significant, task of renormalizing the
vacuum energy for a $\lambda\phi^4$ theory obtained by setting
$h=g=0$ and taking the background scalar field to vanish
($\bv=0$). With this simplification, we find
\begin{eqnarray}
\Gamma^{(2)}&=&\ell^{-\epsilon}\int d^{D+1}x\left\lbrace
\frac{\lambda}{8}G_v^2(x,x)+\frac{1}{2} \delta m^{2(1)}G_v(x,x)
\right\rbrace\nonumber\\
&&+\ell^{-\epsilon}\int\limits_{y=a}^{}d^Dx\;\frac{1}{2}\delta
v^{(1)}G_v(x,x) \;.\label{R24}
\end{eqnarray}

We have already described how to extract the pole part of
$G_v(x,x)$ in Sec.~\ref{oneloop}. At two-loop order we have the
one-loop counterterms $\delta m^{2(1)}$ and $\delta v^{(1)}$
multiplying $G_v(x,x)$, resulting in poles that involve the finite
part of $G_v(x,x)$. Such poles, should they not cancel, will
involve complicated non-local expressions and would render the
theory non-renormalizable with local counterterms. A necessary
part of the analysis will be to show that all such non-local poles
cancel between the three terms of $\Gamma^{(2)}$ given in
(\ref{R24}).

We have given $G_v(x,x)$ in (\ref{A17}) with $G_v(p;y,y)$ given in
(\ref{A16}) for the case of periodic boundary conditions. Because
our main focus is on how the presence of a $\delta$-function
potential changes the renormalization calculation, in order not to
complicate the analysis unnecessarily, we will let
$L\rightarrow\infty$. (The presence of a finite $L$ leads to more
complicated non-local divergences as known in the absence of
$\delta$-function potentials from \cite{DJTS1,BirrellFord}. It
is possible to show that in this more complicated setting all
non-local divergences still cancel as we describe below for
infinite $L$, although we omit the details of this here for simplicity.) With the limit $L\rightarrow\infty$ taken in
(\ref{A16}) we have
\begin{equation}\label{R25}
G_v(p;y,y)=\frac{1}{2\omega_p}-\frac{v}{4\omega_p^2}
\left(1+\frac{v}{2\omega_p}\right)^{-1} e^{-2|y-a|\omega_p}\;.
\end{equation}
This is the result that we would have found had we simply chosen
not to adopt periodic boundary conditions over a finite interval
and taken the whole real line instead.

The evaluation of the Green function expressions that enter the
two-loop part of the effective action in (\ref{R24}) is described
in Appendix~\ref{integrals}. The first two terms, that contain an
integral over $y$ as well as over ${\mathbf x}_\perp$, give rise
to bulk counterterms (ones that multiply $LV_3$) as well as
surface counterterms (ones that multiply $V_3$ but that are
independent of $L$). Because the one-loop counterterms $\delta
m^{2(1)}$ and $\delta v^{(1)}$ multiply $G_v(x,x)$, we must
include the finite part of $G_v(x,x)$ in order to calculate the
complete pole terms of the effective action. From (\ref{B2}), with
$D=3+\epsilon$, we have
\begin{eqnarray}
I(1)&=&\frac{m^2}{4\pi^2\epsilon}+\frac{m^2}{8\pi^2}\left\lbrack
\ln\left(\frac{m^2}{4\pi}\right)+\gamma-1\right\rbrack+\cdots\;,\label{R26a}\\
I(2)&=&-\frac{m}{4\pi}+\cdots\;,\label{R26b}\\
I(3)&=&-\frac{1}{2\pi^2\epsilon}-\frac{1}{4\pi^2}\left\lbrack
\ln\left(\frac{m^2}{4\pi}\right)+\gamma\right\rbrack+\cdots\;,\label{R26c}
\end{eqnarray}
upon expansion about $\epsilon=0$. (The ellipsis indicates terms of order $\epsilon$ that cannot lead to poles in the effective action, although they may contribute to the finite part.)

If we temporarily ignore the surface term in $\Gamma^{(2)}$, and
use the expression found earlier for $\delta m^{2(1)}$ given in
(\ref{R23c}), it is possible to show that
\begin{eqnarray}
\Gamma^{(2)}/V_3&=&-\frac{\lambda m^4}{512\pi^4\epsilon^2}L+
\frac{\lambda mv^2}{512\pi^3\epsilon} -\frac{\lambda vm^2}{512\pi^4\epsilon}\nonumber\\
&&-\frac{\lambda v^3}{1024\pi^4\epsilon^2} +\left( \frac{\lambda
vm^2}{512\pi^4\epsilon}-\frac{\lambda v^3}{1024\pi^4\epsilon}
\right) \ln\left(\frac{m^2\ell^2}{4\pi}\right)\nonumber\\
&&-\frac{\lambda
v^4}{512\pi^2\epsilon}K_4(v)+\ell^{-\epsilon}\int\limits_{y=a}^{}d^Dx\;\frac{1}{2}\delta
v^{(1)}G_v(x,x)\label{R27}
\end{eqnarray}
contains the pole part of the two-loop effective action. The bulk
term, the first term proportional to $L$, involves only a local
divergence that can be dealt with by a local counterterm. In this
bulk term, all of the dependence on $\gamma$ and $\ln(m^2\ell^2)$
that occur at intermediate stages of the calculation has
cancelled, but there are observed to be divergences that involve
$\ln(m^2\ell^2)$ in the surface term. In addition, there is a
non-local divergence proportional to the integral $K_4(v)$
present.

The surface contribution (the last term on the right hand side of
(\ref{R27})) involves
\begin{eqnarray}
\frac{1}{2}\ell^{-\epsilon}\delta v^{(1)}\left.
G_v(x,x)\right|_{y=a} &=&-\frac{\lambda
vm^2}{512\pi^4\epsilon^2}-\frac{\lambda
vm^2}{512\pi^3\epsilon}\nonumber\\
&&+\frac{\lambda v^3}{512\pi^4\epsilon^2} +\left( \frac{\lambda
v^3}{1024\pi^4\epsilon} -\frac{\lambda vm^2}{512\pi^4\epsilon}
\right)\ln\left(\frac{m^2\ell^2}{4\pi}\right)\nonumber\\
&&+\frac{\lambda v^4}{512\pi^2\epsilon}K_4(v)+\cdots\;,\label{R28}
\end{eqnarray}
if we use (\ref{R23f}) for $\delta v^{(1)}$ and (\ref{B6}). If we
use the result found in (\ref{R28}) back in (\ref{R27}) it can be
seen that the non-local divergences that involve $K_4(v)$ cancel,
and that the dependence on $\ln(m^2\ell^2)$ also cancels. We are
left with the simple expression
\begin{equation}\label{R29}
\Gamma^{(2)}/V_3= -\frac{\lambda m^4}{512\pi^4\epsilon^2}L+
\frac{\lambda v^3}{1024\pi^4\epsilon^2} -\frac{\lambda
vm^2}{256\pi^4\epsilon^2}+\cdots
\end{equation}
as the pole part of $\Gamma^{(2)}$. The first term on the right
hand side represents the bulk divergence, present even when there
is no $\delta$-function potential, and the other two terms
represent the poles on the surface. Note that the presence of the
contribution from the one-loop surface counterterm, that involved
the renormalized strength of the $\delta$-function, was crucial in
obtaining a cancellation of the non-local divergences.

The divergences present in (\ref{R29}) can be seen to be removed
by the choice of two-loop vacuum energy counterterm
\begin{equation}\label{R30}
\delta\Lambda^{(2)}=\frac{\lambda m^4}{512\pi^4\epsilon^2}\;,
\end{equation}
and the renormalized surface density
\begin{equation}\label{R31}
\delta\sigma^{(2)}=\frac{\lambda v}{256\pi^4\epsilon^2}
\left(\frac{v^2}{4}-m^2\right)\;.
\end{equation}
This presents a complete proof that to two-loop order the vacuum energy density is renormalizable, and we have computed the necessary counterterms to do this.

\section{\label{vacuum} Vacuum energy}
\setcounter{equation}{0}

In this section we will compute the vacuum energy density to one-loop order for a complex scalar field $\Phi$ in a $\delta$-function potential. We will allow the boundary condition on the field to depend on an arbitrary change of phase around the circle:
\begin{equation}\label{3.1}
\Phi(-L/2)=e^{i2\pi \alpha}\Phi(L/2)\;.
\end{equation}
Here $0\le \alpha\le1$ is the arbitrary phase factor. The case $\alpha=0$ corresponds to periodic boundary conditions, while $\alpha=1/2$ corresponds to antiperiodic boundary conditions. We will look at how $\alpha$ enters the expression for the vacuum energy density.

Similar calculations without a $\delta$-function potential present were performed some time ago. Of special interest is the paper of Ford~\cite{FordPRD}, where it was remarked that the situation is the same as coupling a scalar field to a constant gauge potential, with the case of $\alpha=0$ corresponding to a vanishing gauge field unstable for spinor fields. This was later generalized to non-Abelian gauge theories~\cite{TomsPLB}, and Hosotani~\cite{Hosotani} showed that by computing an effective potential as a function of $\alpha$ it was possible to break non-Abelian gauge symmetries. More recent work \cite{Oda} has studied this mechanism in the brane-world setting where $\delta$-function potentials arise naturally.

The generalization of (\ref{R14}) to the complex scalar field case is obtained simply by multiplying by two (to account for the fact that a complex field can be represented by two real fields) and changing the boundary conditions on the Green function to reflect (\ref{3.1}):
\begin{equation}\label{3.2}
\frac{\partial}{\partial m^2} \Gamma^{(1)}=V_D\int\limits_{-L/2}^{L/2}dy\int\frac{d^Dp}{(2\pi)^D}
\;G_v(p;y,y)\;.
\end{equation}
From (\ref{A12}) we find
\begin{equation}\label{3.3}
G_v(p;y,y)=G_0(p;y,y)-\frac{v G_0(p;y,a)G_0(p;a,y)}{1+vG_0(p;a,a)}
\end{equation}
where $G_0(p;y,y')$ is given by (\ref{A13}) with
\begin{equation}
\sigma_j=\frac{2\pi}{L}(j+\alpha)\label{3.5}
\end{equation}
and $\omega_p^2=p^2+m^2$ as before. From (\ref{A13}) it is easy to see that
\begin{eqnarray}
\int\limits_{-L/2}^{L/2}dy\,G_0(p;y,a)G_0(p;a,y)&=&\sum_{j=-\infty}^{\infty}\frac{1}{L}(\sigma_j^2+\omega_p^2)^{-2}\nonumber\\
&=&-\frac{\partial}{\partial m^2} \sum_{j=-\infty}^{\infty}\frac{1}{L}(\sigma_j^2+\omega_p^2)^{-1}\;.\label{3.6}
\end{eqnarray}
This leads to
\begin{equation}\label{3.7}
\Gamma^{(1)}=\Gamma^{(1)}_{v=0}+\Gamma^{(1)}_{\rm surf}\;,
\end{equation}
where
\begin{equation}\label{3.8}
\Gamma^{(1)}_{v=0}=V_D\sum_{j=-\infty}^{\infty}\int\frac{d^Dp}{(2\pi)^D}\ln\left\lbrack\ell^2(\sigma_j^2+\omega_p^2)\right\rbrack
\end{equation}
is the term present in the case of $v=0$, corresponding to the absence of a $\delta$-function potential, and
\begin{equation}\label{3.9}
\Gamma^{(1)}_{\rm surf}=V_D\int\frac{d^Dp}{(2\pi)^D}\ln\left\lbrack 1+\frac{v}{L}\sum_{j=-\infty}^{\infty}(\sigma_j^2+\omega_p^2)^{-1}\right\rbrack
\end{equation}
is the `surface' contribution to the effective action coming from the $\delta$-function potential.

The sum over $j$ in (\ref{3.9}) is easily evaluated by contour integral methods \cite{Bromwich} to give
\begin{equation}\label{3.10}
\sum_{j=-\infty}^{\infty}\frac{1}{L}(\sigma_j^2+\omega_p^2)^{-1} = \frac{1}{2\omega_p}\,\frac{\sinh(L\omega_p)}{\cosh(L\omega_p)-\cos(2\pi \alpha)}\;.
\end{equation}
The pole part of $\Gamma^{(1)}_{\rm surf}$ as $D\rightarrow3$ can be analyzed as described in Sec.~\ref{renorm} above.

The $\Gamma^{(1)}_{v=0}$ term can be related to the function evaluated first by Ford~\cite{FordPRD}. (See \cite{TomsSAP} for a textbook treatment.) If we define
\begin{equation}\label{3.11}
F(\lambda;\alpha,b)= \sum_{n=-\infty}^{\infty} \left\lbrack(n+\alpha)^2+b^2\right\rbrack^{-\lambda}\;,
\end{equation}
then after suitable analytic continuation
\begin{equation}\label{3.12}
F(\lambda;\alpha,b)=\pi^{1/2}\frac{\Gamma(\lambda-1/2)}{\Gamma(\lambda)} (b^2)^{1/2-\lambda} +f_{\lambda}(\alpha,b)
\end{equation}
where
\begin{equation}\label{3.13}
f_{\lambda}(\alpha,b)=4\sin(\pi\lambda)\int\limits_{b}^{\infty}dx\,(x^2-b^2)^{-\lambda} \Re\left\lbrack e^{2\pi(x+i\alpha)}-1\right\rbrack^{-1}\;.
\end{equation}
It is straightforward to show that
\begin{equation}\label{3.14}
\Gamma^{(1)}_{v=0}=-V_D\ell^{D-3}(4\pi)^{-D/2} \Gamma\left(-\frac{D}{2}\right)\left(\frac{2\pi}{L}\right)^D \,F\left(-\frac{D}{2};\alpha,\frac{mL}{2\pi} \right),
\end{equation}
and that poles come only from the first term on the right hand side of (\ref{3.12}) when used in (\ref{3.14}).

A full renormalization calculation can be performed as we have described in Sec.~\ref{renorm}. However, our aim in this section is to study the role that $\alpha$ and $v$ play in the expression for the vacuum energy density. To do this we will simply compare the vacuum energy for general $\alpha$ with that for $\alpha=0$. (A detailed analysis of the $\alpha=0$ case was given in \cite{TomsPLBdelta}  and will not be repeated here.) Because $\Gamma^{(1)}=V_DL\rho$ with $\rho$ the vacuum energy density, or effective potential, we have
\begin{equation}\label{3.15}
\rot=\rho(\alpha)-\rho(0)
\end{equation}
as the difference between the $\alpha\ne0$ and $\alpha=0$ cases. A bit of calculation shows that
\begin{eqnarray}\label{3.16}
2\pi^2L^4\rot&=&\int\limits_{mL}^{\infty}dy\,y(y^2-m^2L^2)^{1/2}\nonumber\\
&&\times\ln \left\lbrace \frac{y+vL/2 -2ye^{-y}\cos(2\pi\alpha) +(y-vL/2)e^{-2y}}{(1-e^{-y})\lbrack y+vL/2-(y-vL/2)e^{-y} \rbrack}\right\rbrace.
\end{eqnarray}
This same result can be found by the more tedious and lengthy process of expanding about the pole at $D=3$, removing the poles with counterterms as described in Sec.~\ref{renorm}, or else by adopting $\zeta$-function regularization where no poles occur~\cite{TomsPLBdelta}.

\begin{figure}[ht]
\includegraphics{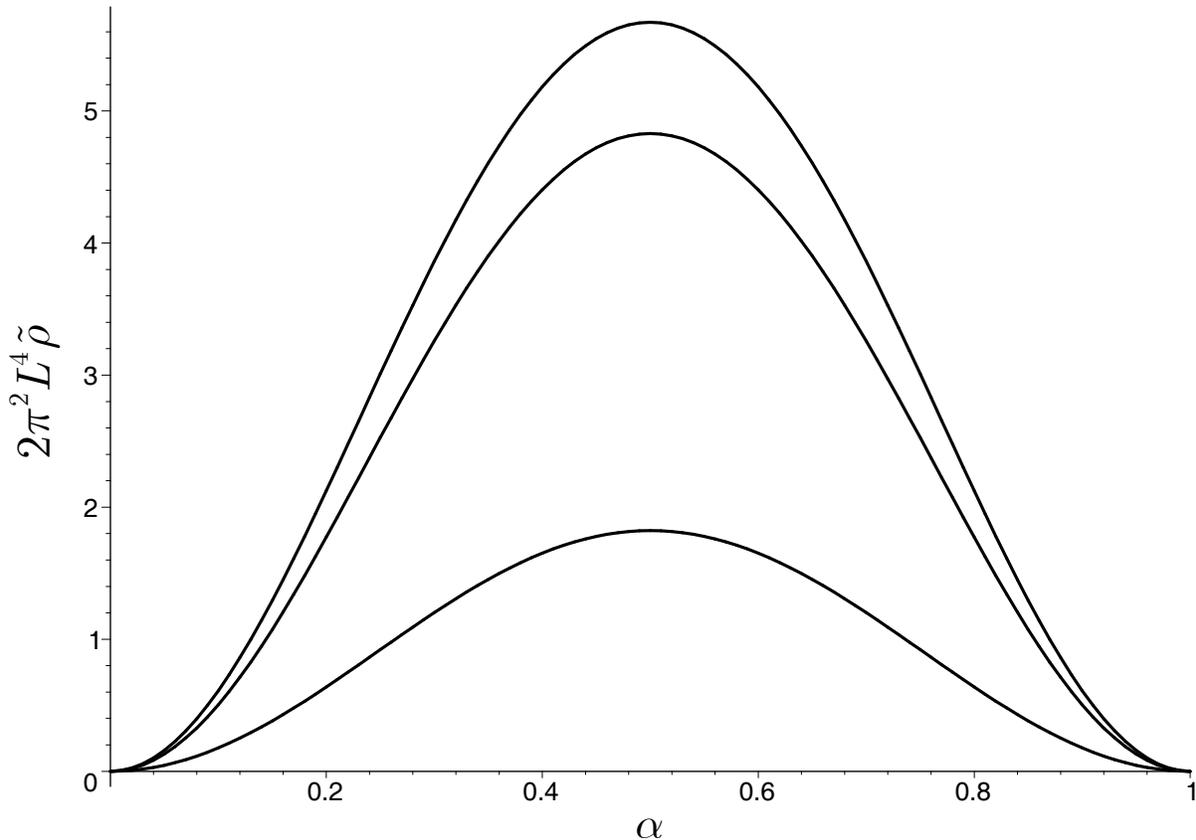}
\caption{This shows plots of $2\pi^2L^4\tilde{\rho}$ as a function of $\alpha$, the parameter that determines the boundary conditions, for three different values of $mL$. The top curve is the result for $m=0$. The middle and lower curves are the results for $mL=1$ and $mL=3$ respectively. In all cases the maximum occurs at $\alpha=1/2$ as the analytic proof described in the text shows. The energy density decays exponentially with $mL$.\label{fig1}}
\end{figure}

If we view $\rot$ as a function of $\alpha$, it is easy to show from (\ref{3.16}) that for $0\le \alpha\le1$, $\rot$ has a global minimum at $\alpha=0$ (or $\alpha=1$) and a global maximum at $\alpha=1/2$. Thus, the case of antiperiodic boundary conditions leads to the maximum energy density, a conclusion that is the same as in the absence of a $\delta$-function potential~\cite{DeWittHartIsham,FordPRD} although the actual expressions for the energy density differ of course. In Fig.~\ref{fig1} we show the result of evaluating (\ref{3.16}) numerically for different values of $mL$, but with $vL$ kept fixed. The vacuum energy decays exponentially with $mL$ exactly as in the case with no $\delta$-function potential present~\cite{TomsPLB}.

We can also study what happens if the strength of the $\delta$-function potential $v$ is varied. For simplicity we will set $m=0$ and $\alpha=1/2$. The result is plotted in Fig.~\ref{fig2} as the solid curve. If we take $vL\gg1$, it is possible to obtain the following asymptotic expansion for $\rot$ in (\ref{3.16}) for $m=0$ but $\alpha$ general:
\begin{eqnarray}
\rot&=&\frac{\pi^2\sin^2(\pi\alpha)}{4vL^5} \Big\lbrace 1-\frac{4}{15vL}\left\lbrack 29-\cos(2\pi\alpha)\right\rbrack\nonumber\\
&&+\frac{2}{3v^2L^2}\left\lbrack 86+3\pi^2+2(8-\pi^2)\cos(2\pi\alpha) +(10-\pi^2) \cos(4\pi\alpha) \right\rbrack\nonumber\\
&&\qquad+\cdots\Big\rbrace.\label{3.17}
\end{eqnarray}
For the case of $\alpha=1/2$ this is plotted as the dotted line in Fig.~\ref{fig2}.

\begin{figure}[ht]
\includegraphics[scale=0.9]{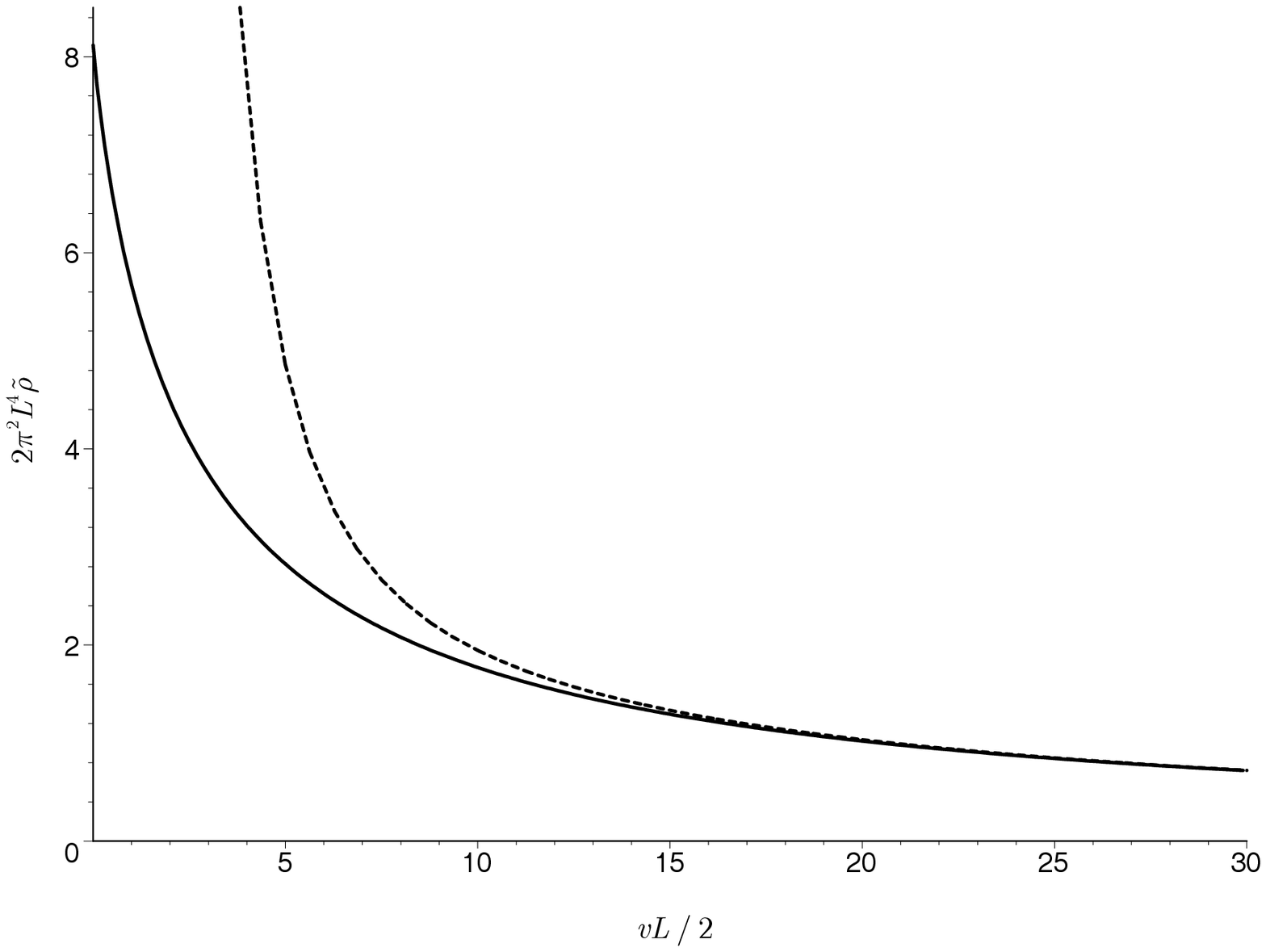}
\caption{The solid line shows the result of $2\pi^2L^4\rot$ as a function of $vL/2$ found from (\ref{3.16}) in the case $m=0$ and $\alpha=1/2$. The dotted line shows the same result found using the asymptotic expansion in (\ref{3.17}). The agreement between the analytic and numerical results become very good once $vL$ is sufficiently large.\label{fig2}}
\end{figure}

We can also analyze the case where $vL$ is small and obtain a reliable asymptotic expansion. The analysis is reasonably involved, so we will simply quote the result. If $vL/2\ll1$ we find from (\ref{3.16}) that (again taking $m=0$ and $\alpha=1/2$ as an example)
\begin{equation}\label{3.18}
\rot\simeq\frac{\pi^4}{12}-\frac{\pi^2}{4}vL+\frac{\pi}{3}(vL)^{3/2}-\beta(vL)^2+\cdots\;,
\end{equation}
where $\beta\simeq0.0288734$ is the result of a numerical evaluation of a simple integral of the exponential integral form. Contributions of the next order in (\ref{3.18}) can also be found and involve non-analytical terms of order $(vL)^{5/2}$ and $(vL)^3\ln(vL)$. The fact that $\rot$ is not analytical at $v=0$ is what complicates attempts to calculate the asymptotic expansion. Similar non-analyticity has been seen in the earlier calculations of \cite{MiltonJPA37} in similar settings and indicates the futility of trying to use a naive perturbative approach about $v=0$ (corresponding to treating the $\delta$-function potential as a perturbative interaction.) To demonstrate the utility of (\ref{3.18}) we plot the approximation (shown as the dotted line) against a numerical evaluation of the exact result (shown as the solid line) in Fig.~\ref{fig3}.
\begin{figure}[htb]
\includegraphics{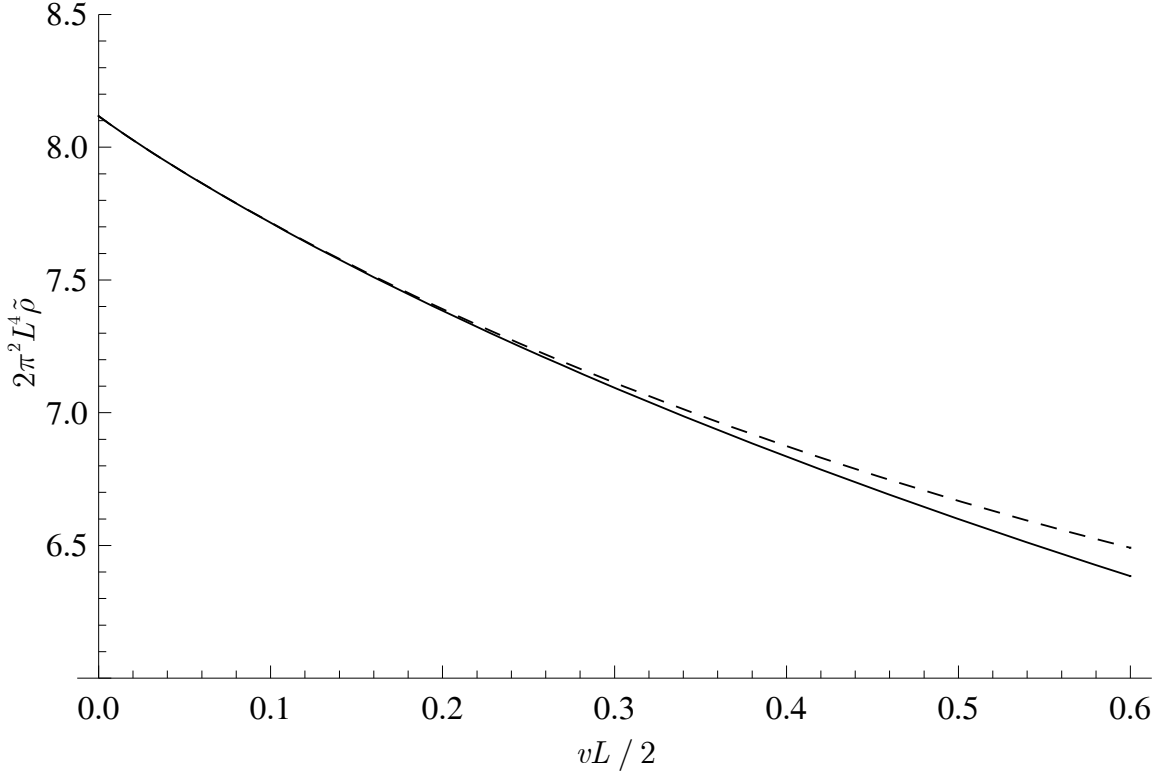}
\caption{The solid line shows the result of $2\pi^2L^4\rot$ as a function of $vL/2$ found from (\ref{3.16}) in the case $m=0$ and $\alpha=1/2$. The dotted line shows the same result found using the asymptotic expansion in (\ref{3.18}). Note that the range is shown over $6-8.5$ rather than extending to 0 to exaggerate the difference between the two curves. As $vL$ decreases, the agreement becomes excellent.\label{fig3}}
\end{figure}

\section{\label{discuss} Discussion and conclusions}
\setcounter{equation}{0}

We have considered the case of an interacting field theory in a non-simply connected spacetime in the presence of a $\delta$-function potential. The necessity for a proper inclusion of boundary interaction terms for renormalizability was discussed. The required counterterms were computed, using dimensional regularization, to one-loop order, and some consideration was given to the two-loop counterterms. It was shown why the complete two-loop calculation was difficult. We then showed how to obtain the effective potential for a complex scalar field with general boundary conditions around a compact spatial dimension, generalizing earlier such studies. A number of approximations were obtained using analytical methods and compared with a numerical evaluation of the exact result for the effective potential in certain cases. In particular, the case of a weakly coupled $\delta$-function potential was shown to result in a non-analytic expression for the vacuum energy that will not show up if a normal weak-field perturbative approach is used.

There are a number of future directions that are worthy of attention. The first is to find a method to obtain the complete set of boundary counterterms to two-loop order, and if possible to proceed beyond two-loop order. In particular it would be of interest to see if the non-analyticity seen in the $\delta$-function vacuum energy could affect the renormalization procedure at higher orders. It would also be of interest to examine how the analysis presented here is modified if there is more than one $\delta$-function present, or if the $\delta$-functions are more than one-dimensional. The case of spherical $\delta$-functions relevant for spherical or cylindrical boundary problems is also of some interest. Having a more complete analysis of the counterterms would enable a renormalization group study to be performed and further illustrate the role of the surface divergences. Finally we mention that it is of interest to examine the complete stress-energy tensor for the interacting case, complementing the free field cases that have been done. Some work has been done on this~\cite{Sam}.

\acknowledgments{I am grateful to Klaus Kirsten for sharing his knowledge of interacting fields on manifolds with boundary.}

\appendix

\section{\label{Green}Green functions}

In this appendix we describe a method for calculating the Feynman
Green function in the presence of one-dimensional
$\delta$-function potentials that generalizes \cite{Solod}. An earlier evaluation of the one-dimensional Green function using an entirely different approach was given by \cite{Blinder}, and by \cite{HR} for scalars and spinors. Take a $(D+1)$-dimensional spacetime
with spacetime coordinates $x^\mu=(\mathbf{x}_\perp,y)$ and adopt
a Euclidean metric. Here $y$ is used to distinguish the coordinate
that enters the $\delta$-functions. Take the potential to be
\begin{equation}
V(y)=\sum_{i=1}^{n}v_i\delta(y,a_i)\;,\label{A1}
\end{equation}
where the $v_i$ are constants and $a_i$ give the locations of the
$\delta$-function singularities. In the case where $y$ runs over
the finite range $\lbrack-L/2,L/2\rbrack$ with $y=-L/2$ identified
with $y=L/2$, we assume that $-L/2<a_i<L/2$ for all
$i=1,\ldots,n$.

We now wish to solve for the Green function $G_v(x,x')$ defined as
the fundamental solution to
\begin{equation}
\left(-\Box+m^2+V(y)\right)G_v(x,x')=\delta(x,x') \;.\label{A2}
\end{equation}
Write
\begin{equation}
G_v(x,x')=\int\frac{d^Dp}{(2\pi)^D}\;
e^{ip\cdot(\mathbf{x}_\perp-\mathbf{x}^{\prime}_\perp)}G_v(p;y,y') \;.\label{A3}
\end{equation}
It then follows from (\ref{A2}) that
\begin{equation}
\left\lbrack -\frac{\partial^2}{\partial y^2}+
\omega_p^2+V(y)\right\rbrack G_v(p;y,y') =\delta(y,y') \;,\label{A4}
\end{equation}
where
\begin{equation}
\omega_p=\left(p^2+m^2\right)^{1/2}\;.\label{A5}
\end{equation}
With $V(y)$ given as in (\ref{A1}) we can use the
$\delta$-functions to set $y=a_i$ in $G_v(p;y,y')$, thereby
obtaining
\begin{equation}
\left\lbrack -\frac{\partial^2}{\partial y^2}
+\omega_p^2\right\rbrack G_v(p;y,y') =\delta(y,y')
-\sum_{i=1}^{n}v_i\,\delta(y,a_i)G_v(p;a_i,y') \;.\label{A6}
\end{equation}

Now define $G_0(p;y,y')$ to be the solution to (\ref{A6}) with
$v_i=0$; that is, for no $\delta$-functions present. This means
that
\begin{equation}
\left\lbrack -\frac{\partial^2}{\partial y^2}+
\omega_p^2\right\rbrack G_0(p;y,y') =\delta(y,y') \;.\label{A7}
\end{equation}
We can set $y'=a_i$ in (\ref{A7}) and then use this to eliminate
$\delta(y,a_i)$ in (\ref{A6}). Rearranging the result gives
\begin{eqnarray}
&&\left\lbrack -\frac{\partial^2}{\partial y^2}+\omega_p^2\right\rbrack
\left\lbrace G_v(p;y,y') +\sum_{i=1}^{n}v_i\,G_0(p;y,a_i)G_v(p;a_i,y')
\right\rbrace\nonumber\\
&&\hspace{75mm} =\delta(y,y') \;.\label{A8}
\end{eqnarray}
Since the solution to (\ref{A7}) should be unique (given the
boundary conditions) we may identify the expression in braces in
(\ref{A8}) with $G_0(p;y,y')$. This gives us
\begin{equation}
G_v(p;y,y') +\sum_{i=1}^{n}v_i\,G_0(p;y,a_i)G_v(p;a_i,y') =G_0(p;y,y') \;.\label{A9}
\end{equation}
If we set $y=a_j$ in (\ref{A9}) we find
\begin{equation}
\sum_{i=1}^{n}\left\lbrace \delta_{ij}+v_iG_0(p;a_j,a_i)\right\rbrace G_v(p;a_i,y') =G_0(p;a_j,y') \;.\label{A10}
\end{equation}
This gives us a set of equations that determines $G_v(p;a_i,y')$
that occurs in (\ref{A9}) in terms of $G_0(p;y,y')$. Thus we
obtain $G_v(p;y,y')$ from a knowledge of the Green function in the
absence of a $\delta$-function potential.

In the simplest case of a single $\delta$-function potential
($n=1$ in (\ref{A10})) it is easy to see that
\begin{equation}
G_v(p;a,y')=\frac{G_0(p;a,y')}{1+vG_0(p;a,a)} \;.\label{A11}
\end{equation}
This gives us
\begin{equation}
G_v(p;y,y')=G_0(p;y,y')-  \frac{v\; G_0(p;y,a)G_0(p;a,y')}{1+vG_0(p;a,a)} \;,\label{A12}
\end{equation}
if we use (\ref{A9}). The case of more than one $\delta$-function
potential can be dealt with in a similar manner, although of
course the details become more involved. For example, it is
straightforward to recover the Green function for two $\delta$
functions used by Milton~\cite{MiltonJPA37}.

For the present paper we are concerned with a single
$\delta$-function, and assume $-L/2\le y\le L/2$ with the
endpoints identified. For the case of periodic boundary conditions
we have
\begin{equation}
G_0(p;y,y')=\sum_{j=-\infty}^{\infty}\frac{1}{L}e^{i\sigma_j(y-y')} \left(\sigma_j^2+\omega_p^2\right)^{-1} \;,\label{A13}
\end{equation}
where
\begin{equation}
\sigma_j=\frac{2\pi j}{L}\;.\label{A14}
\end{equation}
Note that $G_0(p;y,y')$ can only depend on $|y-y'|$, a result that
is easily seen by relabelling $j$ to $-j$ in the sum. The sum over
$j$ can be computed using contour integral methods to
give~\cite{GR}
\begin{equation}
G_0(p;y,y')=\frac{1}{2\omega_p}\;\frac{\cosh\left(\frac{L}{2}\omega_p-\omega_p|y-y'|\right)}{\sinh\left(\frac{L}{2}\omega_p\right)}  \;,\label{A15}
\end{equation}
This expression is sufficient to determine $G_v(p;y,y')$ and hence
the full Green function in the presence of a $\delta$-function
potential. Note that as $L\rightarrow\infty$ we recover the flat
spacetime result of $G_0(p;y,y')=e^{-\omega_p|y-y'|}/(2\omega_p)$.

We are mainly concerned with the coincidence limit of the Green
function. From (\ref{A12}) and (\ref{A15}) we find
\begin{eqnarray}
G_v(p;y,y)&=&G_0(p;y,y)-  \frac{v\; G_0^2(p;y,a)}{1+vG_0(p;a,a)}\nonumber\\
&&\hspace{-15mm}=\frac{1}{2\omega_p}\coth\left(\frac{L}{2}\omega_p\right)-\frac{v}{4\omega_p^2}\;\frac{\cosh^2\left\lbrack\left(\frac{L}{2}-|y-a|\right)\omega_p\right\rbrack}{\sinh^2\left(\frac{L}{2}\omega_p\right)\left\lbrack1+\frac{v}{2\omega_p}\coth\left(\frac{L}{2}\omega_p\right)\right\rbrack} \;.\label{A16}
\end{eqnarray}
The coincidence limit of the Green function is then obtained from
(\ref{A3}),
\begin{equation}
G_v(x,x)=\int\frac{d^Dp}{(2\pi)^D}G_v(p;y,y) \;,\label{A17}
\end{equation}
with (\ref{A16}) used on the right hand side.

\section{\label{integrals} Evaluation of some integrals}

We define
\begin{equation}
I(\alpha)=\int\frac{d^Dp}{(2\pi)^D}\;\omega_p^{-\alpha} \;,\label{B1}
\end{equation}
where $\omega_p=(p^2+m^2)^{1/2}$. Making use of the standard
integral representation for the $\Gamma$-function \cite{GR} it is easy to
show that
\begin{equation}
I(\alpha)=(4\pi)^{-D/2}\;
\frac{\Gamma\left(\frac{\alpha-D}{2}\right)}{\Gamma
\left(\frac{\alpha}{2}\right)}(m^2)^{(D-\alpha)/2} \;.\label{B2}
\end{equation}
This is a standard result of dimensional regularization~\cite{tHooftVeltman}.

We also define
\begin{equation}
K_n(v)=\int\frac{d^Dp}{(2\pi)^D}\;\omega_p^{-n}\left(1+\frac{v}{2\omega_p}\right)^{-1} \;.\label{B3}
\end{equation}
It is simple to show that $K_n(v)$ satisfies the recursion relation
\begin{equation}
K_n(v)=I(n)-\frac{v}{2}K_{n+1}(v)\;,\label{B4}
\end{equation}
where $I(n)$ is defined as in (\ref{B1}). We are concerned with
the case $D\rightarrow3$ in this paper, and simple power counting
shows that $K_n(v)$ is finite as $D\rightarrow3$ for $n\ge4$. The
recursion relation (\ref{B4}) enables us to isolate the divergent
parts of $K_n(v)$ in terms of the simpler integral given in
(\ref{B1}) and (\ref{B2}).

With these preliminaries over, we can now evaluate the divergent
parts of the Green function expressions that enter the two-loop
vacuum energy. First of all, in the large $L$ limit, from
(\ref{A16}) and (\ref{A17}) we have
\begin{eqnarray}
\int\limits_{-L/2}^{L/2}dy\;G_v(x,x)&=&\int\frac{d^Dp}{(2\pi)^D}
\left\lbrace \frac{L}{2\omega_p} -\frac{v}{4\omega_p^2}
\left(1+\frac{v}{2\omega_p}\right)^{-1}
\int\limits_{-\infty}^{\infty} dy\,e^{-2|y-a|\omega_p}\right\rbrace\nonumber\\
&=&\frac{L}{2}I(1)-\frac{v}{4}\int\frac{d^Dp}{(2\pi)^D}\;
\omega_p^{-2}\left(\omega_p+\frac{v}{2}\right)^{-1}\nonumber\\
&=&\frac{L}{2}I(1)-\frac{v}{4}K_3(v)\nonumber\\
&=&\frac{L}{2}I(1)-\frac{v}{4}I(3)+\frac{v^2}{8}K_4(v) \;.\label{B5}
\end{eqnarray}
In the last line we have used (\ref{B4}). The result in (\ref{B5})
is exact. The first two terms contain poles as $D\rightarrow3$ and
the last term is finite.

We also need
\begin{eqnarray}
\left.G_v(x,x)\right|_{y=a}&=&\int\frac{d^Dp}{(2\pi)^D}
\left\lbrace \frac{1}{2\omega_p} -\frac{v}{4}\omega_p^{-2}
\left(1+\frac{v}{2\omega_p}\right)^{-1}\right\rbrace\nonumber\\
&=&\frac{1}{2}I(1)-\frac{v}{4}K_2(v)\nonumber\\
&=&\frac{1}{2}I(1)-\frac{v}{4}I(2)+\frac{v^2}{8}I(3)-\frac{v^3}{16}K_4(v)\;.\label{B6}
\end{eqnarray}
Any poles can come only from the first three terms on the right hand side as $K_4(v)$ is finite as $D\rightarrow3$. .

Finally, we need
$\displaystyle{\int\limits_{-L/2}^{L/2}dy\,G_v^2(x,x)}$. We can
use (\ref{A16}) and (\ref{A17}) to show in the limit of
$L\rightarrow\infty$ that
\begin{eqnarray}
\int\limits_{-L/2}^{L/2}dy\,G_v^2(x,x)&=&
\int\limits_{-L/2}^{L/2}dy\int
\frac{d^Dp}{(2\pi)^D}\frac{d^Dq}{(2\pi)^D}
\Big\lbrace \frac{1}{4\omega_p\omega_q}\nonumber\\
&&-
\frac{v}{4\omega_p\omega_q^2} \left(1+\frac{v}{2\omega_q}\right)^{-1}e^{-2|y-a|\omega_q}\nonumber\\ &&\hspace{-10mm}+\frac{v^2}{16}\omega_p^{-2}\omega_q^{-2}
\left(1+\frac{v}{2\omega_q}\right)^{-1}
\left(1+\frac{v}{2\omega_p}\right)^{-1}
e^{-2|y-a|(\omega_p+\omega_q)}\Big\rbrace\nonumber\\
&=&\frac{L}{4}I^2(1)-\frac{v}{4}I(1)K_3(v)+J\;,\label{B7}
\end{eqnarray}
with
\begin{eqnarray}
J&=&\frac{v^2}{16} \int\frac{d^Dp}{(2\pi)^D}
\frac{d^Dq}{(2\pi)^D} \omega_p^{-2}\omega_q^{-2}\nonumber\\
&&\times\left(1+\frac{v}{2\omega_q}\right)^{-1}
\left(1+\frac{v}{2\omega_p}\right)^{-1}(\omega_p+\omega_q)^{-1} \;.\label{B8}
\end{eqnarray}
The only complication is the evaluation of $J$, because the double
integral does not factorize. If we consider the integral over $q$
first it is easily seen that
\begin{eqnarray}
\int\frac{d^Dq}{(2\pi)^D}\omega_q^{-2}\left(1+\frac{v}{2\omega_q}\right)^{-1}
(\omega_p+\omega_q)^{-1}&=& \frac{2}{v}\omega_p^{-1}I(1)\nonumber\\
&&\hspace{-25mm}-
\frac{2}{v}\left(\omega_p-\frac{v}{2}\right)^{-1}K_1(v)\nonumber\\
&&\hspace{-25mm}+\omega_p^{-1}\left(\omega_p-\frac{v}{2}\right)^{-1}
\int\frac{d^Dq}{(2\pi)^D}(\omega_p+\omega_q)^{-1}\;.\label{B9}
\end{eqnarray}
If we use (\ref{B9}) in (\ref{B8}), we find that
\begin{equation}
J=\frac{v}{8}I(1)K_3(v)-\frac{1}{4v}K_1(v)\left\lbrack K_1(v)+K_1(-v)-2I(1)\right\rbrack+J_1 \label{B10}
\end{equation}
where
\begin{equation}
J_1=\frac{v^2}{16}\int\frac{d^Dp}{(2\pi)^D}
\omega_p^{-2}\left(\omega_p^2-\frac{v^2}{4}\right)^{-1}F_1(p)\;,\label{B11}
\end{equation}
with
\begin{equation}
F_1(p)=\int\frac{d^Dq}{(2\pi)^D}(\omega_q+\omega_p)^{-1}\;.\label{B12}
\end{equation}
We only require the pole part of $J$ for the divergent part of the
effective action. It is easy to show that
\begin{equation}
F_1(p)=I(1)-\omega_pI(2)+\omega_p^2I(3)-\omega_p^3F_2(p) \label{B13}
\end{equation}
with
\begin{equation}
F_2(p)=\int\frac{d^Dq}{(2\pi)^D}\omega_q^{-3}(\omega_q+\omega_p)^{-1}\;.\label{B14}
\end{equation}
$F_2(p)$ is finite as $D\rightarrow3$, and in setting $D=3$ we find
\begin{eqnarray}
F_2(p)&=&\frac{m}{4\pi}\omega_p^{-2}-\frac{1}{2\pi^2}
\omega_p^{-1}\nonumber\\
&&+\frac{m}{2\pi^2}\omega_p^{-2}
\left(\frac{\omega_p^2}{m^2}-1\right)^{1/2}
\ln\left\lbrack \frac{\omega_p}{m}+
\left(\frac{\omega_p^2}{m^2}-1\right)^{1/2}\right\rbrack \;.\label{B15}
\end{eqnarray}

Using (\ref{B13}) and (\ref{B15}) in (\ref{B11}) results in
\begin{eqnarray}
J_1&=&\frac{v^2}{32}I(1)\left\lbrack K_4(v)+K_4(-v)
\right\rbrack +\frac{v^3}{64}I(2) \left\lbrack K_4(v)-K_4(-v)\right\rbrack\nonumber\\
&&+\frac{v^4}{128}I(3)\left\lbrack K_4(v)+K_4(-v)
\right\rbrack-\frac{mv^2}{64\pi}I(3)+\frac{v^2}{32\pi^2}I(2)\nonumber\\
&&+\frac{mv^3}{256\pi}\left\lbrack K_4(v)-K_4(-v)
\right\rbrack+\frac{v^4}{256\pi^2}\left\lbrack K_4(v)+K_4(-v)\right\rbrack\nonumber\\
&&-\frac{mv^2}{32\pi^2}\int\frac{d^Dp}{(2\pi)^D}
\omega_p^{-1}\left(\omega_p^2-\frac{v^2}{4}\right)^{-1}
\left(\frac{\omega_p^2}{m^2}-1\right)^{1/2}\nonumber\\
&&\qquad\qquad\times
\ln\left\lbrack \frac{\omega_p}{m}+
\left(\frac{\omega_p^2}{m^2}-1\right)^{1/2}\right\rbrack \;.\label{B16}
\end{eqnarray}
This result is exact. We are only after the pole part of $J$, and
hence $J_1$, so we can drop all terms that are finite as
$D\rightarrow3$. Of the expressions $I(1),I(2)$ and $I(3)$, only
$I(1)$ and $I(3)$ contain poles as $D\rightarrow3$. $K_4(\pm v)$
is finite as $D\rightarrow3$. The only remaining question concerns
the last term in (\ref{B16}) as $D\rightarrow3$. To see the pole
structure of this integral, expand the integrand in powers of
$\omega_p$. Terms that fall off with $\omega_p$ faster than
$\omega_p^{-4}$ will converge as $D\rightarrow3$. We find
\begin{eqnarray}
m\int\frac{d^Dp}{(2\pi)^D}\omega_p^{-1}
\left(\omega_p^2-\frac{v^2}{4}\right)^{-1}
\left(\frac{\omega_p^2}{m^2}-1\right)^{1/2}
\ln\left\lbrack \frac{\omega_p}{m}+
\left(\frac{\omega_p^2}{m^2}-1\right)^{1/2}\right\rbrack&&\nonumber\\
&&\hspace{-10cm} \simeq  \int\frac{d^Dp}{(2\pi)^D}
\left\lbrace\omega_p^{-2}\ln(2\omega_p/m)+\cdots\right\rbrace\;,\label{B17}
\end{eqnarray}
where terms that are finite as $D\rightarrow3$ have been dropped.
From (\ref{B1}) it is easy to see that
\begin{equation}
\int\frac{d^Dp}{(2\pi)^D}\omega_p^{-\alpha}\ln\omega_p=-I'(\alpha)\;.\label{B18}
\end{equation}
Because $I(2)$ is finite (after regularization) as
$D\rightarrow3$, so is $I'(2)$. This means that (\ref{B17}) is
finite after regularization as $D\rightarrow3$. We conclude that
the last term of (\ref{B16}) does not contain any pole terms.

Retaining only terms that can contain poles as $D\rightarrow3$, we
find from (\ref{B10}) that
\begin{equation}
J=\frac{v^2}{16}I(2)I(3)-\frac{v^3}{32}I^2(3)
-\frac{mv^2}{64\pi}I(3)+\frac{v^4}{32}I(3)K_4(v)+\cdots\;.\label{B19}
\end{equation}
Liberal use of the recursion relation (\ref{B4}) has been made
here. It is noteworthy that all of the terms that involve $K(-v)$
at intermediate stages of the calculation have cancelled. The
result for (\ref{B7}) becomes
\begin{eqnarray}
\int\limits_{-L/2}^{L/2}dy\;G_v^2(x,x)&=&\frac{L}{4}I^2(1)-
\frac{v}{4}I(1)I(3)+\frac{v^2}{8}I(1)K_4(v)+\frac{v^2}{16}I(2)I(3)\nonumber\\
&&-\frac{v^3}{32}I^2(3)-\frac{mv^2}{64\pi}I(3)+\frac{v^4}{32}I(3)K_4(v)+\cdots \;.\label{B20}
\end{eqnarray}
Again terms that are finite as $D\rightarrow3$ have been dropped.
The presence of non-local pole terms coming from $I(1)$ and $I(3)$
multiplying $K_4(v)$ can be noted. Such terms must cancel for renormalizability with local counterterms, and it is shown in Sec.~\ref{twoloop} that this is the case.

\end{document}